\documentclass[prd,twocolumn,showpacs,preprintnumbers,amsmath,amssymb]{revtex4}
\usepackage{graphicx}\usepackage{dcolumn}
\usepackage{bm}
\usepackage{amssymb}\usepackage{amsmath}\usepackage{amsthm}\usepackage{latexsym}\newcommand{\be}{\begin{equation}}\newcommand{\ee}{\end{equation}}\newcommand{\bea}{\begin{eqnarray}}\newcommand{\eea}{\end{eqnarray}}
\begin{document}
\title{Reply to ``Can gravitational dynamics be obtained by diffeomorphism 
invariance of action?"}
\author{Thomas P.~Sotiriou\footnote[2]{sotiriou@sissa.it} and Stefano Liberati\footnote[3]{liberati@sissa.it}}
\affiliation{SISSA/ISAS, via Beirut 2-4, 34014, Trieste, Italy and INFN, Sezione di Trieste}

\begin{abstract}
In a previous work  we showed 
that, in a suitable setting, one can use diffeomorphism invariance in order to derive gravitational field equations from boundary terms of the gravitational action. Standing by our results we reply here to a recent comment questioning their validity. 
\end{abstract}


\maketitle

In \cite{us} we claimed that, for a general gravitational action that leads to second order field equations, one can use surface terms in order to derive these equations for the case of pure gravity. This procedure is based on the requirement of diffeomorphism invariance but it also requires to adopt a special class of diffemorphisms which then leads to consider surface terms whose boundaries are local Rindler horizons (a notion firstly introduced in \cite{Jacobson}).
In order to avoid being repetitive we are not going to review in detail our analysis  here, nor are we going to state its implications, since these are extensively discussed in \cite{us}. A similar approach has been used in \cite{paddy} (see also references therein) for the Einstein--Hilbert action and later generalized to include the Gauss-Bonnet action in \cite{paddy2}. 

In \cite{gao} our findings are questioned on the base that the so called ``healed Einstein--Hilbert action'' 
\be
\label{b}
{S}^{\prime}_{\rm EH}=\frac{1}{16 \pi G}\left[\int_U d^4 x \sqrt{-g}\, R+2\int_{\partial U} d^3 x \sqrt{h}\, K\right],
\ee
is manifestly diffeomorphism invariant and hence no information can be extracted from it using this a symmetry. Moreover a mathematical derivation is presented aimed at showing that the surface terms arising in our approach are identically closed forms and therefore their integrals over the boundary  vanish identically and not due to the fact that the field equations are satisfied.

Let us first stress that the action (\ref{b}) was given in \cite{us} as part of a review of the standard derivation of the field equation from the Einstein--Hilbert action, through  metric variation. 
Indeed our approach applies to much more general actions ({\it e.g.}~Lovelock type actions), besides those commonly used to derive Einstein's equations.
 Additionally, even if one wants to derive the Einstein field equations, it is well known that (\ref{b}) is not the unique action whose variation will lead to them \cite{lan}.  Actually Einstein himself had noticed very early that the field equations of General Relativity could be derived via the variation of a non-covariant action \cite{einstein1,einstein2}. A typical example of such an action is the following:
\be
\label{scrod}
S_{\rm{G}}^{\prime}=\frac{1}{16\pi G}\int_U d^4 x \sqrt{-g} g^{\mu\nu}\left(\Gamma^{\alpha}_{\phantom{a}\beta\mu} \Gamma^{\beta}_{\phantom{a}\alpha\mu}-\Gamma^{\alpha}_{\phantom{a}\mu\nu}\Gamma^{\beta}_{\phantom{a}\alpha\beta}\right),
\ee
which is sometimes referred to as the Schr\"odinger form \cite{schro}.  Note that the variations of  (\ref{b}) and of  non-covariant actions leading to the same field equations, such as (\ref{scrod}) for example, are identical. Indeed the prescription for building up the so called ``bulk action"' in \cite{us} would naturally lead to  Schr\"odinger type forms for the general gravitational actions considered there, given that to our knowledge the Einstein--Hilbert action can be ``healed'' (modified to have a well defined variation) and still remain covariant only by taking form (\ref{b}). 

Nonetheless, since the authors of \cite{gao} concentrate on action (\ref{b}) and what is claimed is that it is automatically diffeomorphism invariant, being built out of generally covariant scalars, the following comments are due. 

While it is definitely true that action (\ref{b}) appears to be written in a generally covariant form, still it might be non-trivial to understand whether, and under which conditions, such an expression is indeed invariant under diffeomorphisms. 
In fact one can easily write non-covariant expressions in a covariant form by introducing a background geometric structure, as it was pointed out already in 1917 \cite{kretsch}.
This is exactly the case with the second integral in (\ref{b}): Even though it is written in a generally covariant way it is, however, foliation dependent. Therefore, the action (\ref{b}) cannot be considered background independent (which is the actual physical property usually enforced by requiring diffeomorphism invariance) even though it is indeed manifestly covariant (see e.g. the relevant discussion in~\cite{giulini}). 
 See also the appendix of \cite{newpaddy} for an enlightening discussion about the covariance and the foliation dependence of (\ref{b}).

What is more, the results on \cite{us} are not derived by using the standard diffeomorphism invariance, {\it i.e.~}invariance under diffeomorphism whose generator vanishes on the boundary and therefore map a manifold $U$ on itself. On the contrary, what is used is a specific class of diffemorphisms, parallel to the boundary, under which the action does not have to be invariant {\it a priori}, since they do not necessarily map $U$ on itself. These diffeomorphisms end up mapping $U$ on $U$ only after suitable constraints on the generator of the diffeomorphism are imposed.  Such constraints can be generically fulfilled without imposing any restriction on the form of the metric  only by considering regions $U$ whose boundaries $\partial U$ are local Rindler horizons~\footnote{We address the reader to \cite{Jacobson} for a detailed analysis of the general use of the local  Rindler horizons and to \cite{us} for details on the specifics related to our approach.} as in this case one can ask the action to be invariant under the special class of diffeomorphisms whose generating vector is taken to satisfy the Killing equation in a neighborhood of a point P. It must be stressed that this part of our procedure (common also to the approach reported in \cite{paddy,paddy2}) takes advantage of the background manifold structure, since by invoking the equivalence principle the neighborhood of each point is considered to be flat. In this sense it goes well beyond standard diffeomorphism invariance.

Based on the three points made above, we believe that the answer of the main question raised in \cite{gao}, which can be summarized as ``how can one obtain information by applying diffeomorphism invariance on a manifestly covariant action'', comes in a straightforward manner:  

First of all,  the Einstein--Hilbert action (\ref{b}) considered in \cite{gao} is not truly background independent for the above mentioned reasons. Second, one does not need to use action (\ref{b}) for the whole procedure as reported in \cite{us}, in the same way that one does not have to use this specific action in order to derive Einstein's equation through standard metric variation.  In general, if one refrains from introducing background structures, like a preferred foliation, the gravitational action leading to the Einstein equations will have an explicitly non-covariant form like e.g. (\ref{scrod}). 
Finally, and most importantly, our approach picks up a very specific class of diffemorphisms under which even a generally covariant action needs to be invariant only when considering small, non-compact, regions whose boundaries are local Rindler horizons (which can be defined through each point of spacetime thanks to its manifold structure). Requiring an action, covariant or not, to be invariant under this specific class of diffeomorphisms is bound to provide some constraint equation.

In this sense our work does not say that standard diffeomorphism invariance is enough to recover the field equations for pure gravity, it provides instead an operational prescription about how to recover the field equations using the diffeomorphism invariance of the action and the knowledge about the local structure of spacetime.  

Having addressed the main conceptual objection presented in \cite{gao}, we now want to point out the flaw in the mathematical argument used in their work to support their claims.

 As correctly stated in \cite{gao}, the field equations are derived in \cite{us} via the expression
\be
\label{step}
\int_{\cal{H}} d^3 x \sqrt{|h|}\, G_{\mu\nu} \xi^\mu n^\nu=0,\ee
where $\cal{H}$ denote the local Rindler horizon, $\xi^\mu$ the local Killing vector and $n^\nu$ the normal to the horizon (see \cite{us} for explanations regarding notation). Eq.~(\ref{step}) is taken in \cite{us} (and also in other approaches invoking the local Rindler horizon  \cite{paddy,newpaddy}) to imply the vanishing of the integrand and therefore the field equation
\be
\label{fe}
G_{\mu\nu}-\Lambda g_{\mu\nu}=0,
\ee
where $\Lambda$ is an arbitrary (cosmological) constant. These equations will have to hold throughout spacetime since each point can be consider to be part of some local Rinder horizon $\cal{H}$ \cite{us}.

It is claimed in \cite{gao} that $\sqrt{|h|}G_{\mu\nu}\xi^\mu n^\nu$ is a closed form. If this is the case it should also be exact according to \cite{waldp} and consequently its integral over a compact boundary will vanish. Incidentally, let us notice that  in \cite{gao}  it is claimed that this criticism to our work does not apply to \cite{paddy,newpaddy} due to, non specified, essential distinctions in the two derivations. For clarity's sake let us stress that for the case of pure gravity, despite the differences in the two approaches, both in \cite{us} and in \cite{paddy,newpaddy}, one arrives at eq.~(\ref{step}) before deriving the field equations, and therefore, if  $\sqrt{|h|}G_{\mu\nu}\xi^\mu n^\nu$ is indeed a closed form  and the integral in eq.~(\ref{step}) was taken over a compact surface as claimed in \cite{gao} then the criticism presented there should apply also in \cite{paddy,newpaddy}.

In order to show that $\sqrt{|h|}G_{\mu\nu}\xi^\mu n^\nu$ is a closed form the following equation is used in \cite{gao}:
\be
\label{volumes}
d_e(G_{\mu\nu}\xi^\mu n^\nu \tilde{\epsilon}_{bcd})=d_e (\epsilon_{abcd}G^{ak}\xi_k)=(\nabla_a G^{ak}\xi_k) \epsilon_{ebcd},
\ee 
where $\epsilon_{abcd}$ and $\tilde{\epsilon}_{bcd}$ are the volume elements on the region $U$ and its boundary $\partial U$ respectively and $d$ is the derivative operator that maps a $p-1$-form to a $p$-form as defined in \cite{waldp}. It is then claimed that due to the Bianchi identity and the fact that $\xi^\mu$ satisfies the Killing equations one can write
\be
\nabla_a G^{ak}\xi_k=\nabla_a (G^{ak})\xi_k+G^{ak}\nabla_a\xi_k=0,
\ee
and therefore show that $d_e(G_{\mu\nu}\xi^\mu n^\nu \tilde{\epsilon}_{bcd})=0$

However, one has to simply notice that in our case $\xi^\mu$ satisfies the Killing equation only on the boundary, and more precisely on the local Rindler horizon, and not throughout $U$ as the above argument requires. Therefore, one cannot claim that  $\sqrt{|h|}G_{\mu\nu}\xi^\mu n^\nu$ is a closed form based on the argument presented in \cite{gao}. However, it is actually possible to show that $\sqrt{|h|}G_{\mu\nu}\xi^\mu n^\nu$ is a closed form using the constraint equations of General Relativity, without having to impose that $\xi^\mu$ satisfies the Killing equation throughout spacetime, even though this was missed in \cite{gao}. Nonetheless, this is not sufficient for the integral in eq.~(\ref{step}) to vanish identically, since this also requires that this integral is take over a compact surface. This is obviously not the case here or in other approaches that make use of the local Rindler horizon \cite{paddy,newpaddy}, since the integral is over the local Rindler horizon and not over some global boundary.

To clarify this we recall that the local Rindler horizon is defined in the following way \cite{Jacobson}: By invoking the equivalence principle consider a small region around any point P as locally flat and then introduce a Rindler frame. Then the null hypersurface passing from P will act as a local Rindler horizon for a suitable congruence of observers. Of course the local Rindler horizon is a notion of approximate nature defined in an open neighborhood of each point and therefore it cannot constitute a compact surface as required for the integral in eq.~(\ref{step}) to vanish identically. Therefore, equation (\ref{step}) is not a trivial identity as claimed in \cite{gao} and the vanishing of the integral on its left hand side indeed implies that the integrand vanishes leading to eq.~(\ref{fe}).

In conclusion, we believe that the claims presented in \cite{gao} are not well based and should not be considered as an argument against the approach presented in \cite{us} or in other relevant papers such as ref.\cite{paddy,paddy2}.

\section*{Acknowledgements}
The authors wish to thank T.~Padmanabhan for helpful discussions.

\end{document}